\documentclass[showpacs,preprintnumbers,amsmath,amssymb,endfloats*]{revtex4}
\usepackage{graphicx}

\begin{document}
\def \ee {\varepsilon}
\thispagestyle{empty}
\title{
Lifshitz-type formulas for graphene and single-wall
carbon nanotubes: van der Waals and Casimir interactions}

\author{
M.~Bordag, B.~Geyer,
G.~L.~Klimchitskaya,\footnote{
On leave from
North-West Technical University,\\
St.Petersburg, Russia.}
and V.~M.~Mostepanenko\footnote{
On leave from
Noncommercial Partnership  ``Scientific\\ Instruments'', Moscow, Russia.}
}

\affiliation{
Center of Theoretical Studies and Institute for Theoretical Physics,\\
Leipzig University, Augustusplatz 10/11, D-04109 Leipzig, Germany
}

\begin{abstract}
Lifshitz-type formulas are obtained for the van der Waals
and Casimir interaction between graphene and a material plate,
graphene and an atom or a molecule, and between a single-wall
carbon nanotube and a plate. The reflection properties of
electromagnetic oscillations on graphene are governed by the
specific boundary conditions imposed on the infinitely thin
positively charged plasma sheet, carrying a continuous fluid with
some mass and  charge density. The obtained formulas
are applied to graphene interacting with Au and Si plates, to
hydrogen atoms and molecules interacting with graphene, and
to single-wall carbon nanotubes interacting with Au and Si plates.
The generalizations to more complicated carbon nanostructures
are discussed.
\end{abstract}
\maketitle

\section{Introduction}
It has been known that two neutral atoms or molecules separated
by a distance which is rather small but much larger than the atomic
dimensions interact through the van der Waals force. As was shown
by London, the van der Waals force arises in second order
perturbation theory from the dipole-dipole interaction. It is caused
by the dispersions of dipole operators, i.e., by quantum fluctuations
\cite{0}. Being a quantum phenomenon, it depends on the Planck
constant $\hbar$. In consequence of interatomic interactions,
the van der Waals force acts also between an atom and a macroscopic
body and between two closely spaced macroscopic bodies. If the
separation between two atoms, an  atom and a macroscopic body or
between two  macrobodies is sufficiently large, so that the
retardation of the electromagnetic fluctuating interaction
contributes significantly, the van der Waals force depends on
both $\hbar$ and the velocity of light $c$. In this regime
the force is usually labeled as Casimir \cite{01} or (in the case
of atom-atom or atom-wall interaction) Casimir-Polder \cite{02}
force.
Van der Waals and Casimir forces play an important role in the
interaction of a single layer of graphite (hereafter, graphene) with
macroscopic bodies (a material plate or a semispace) and with
microparticles (an atom or a molecule). The physics of interactions
between graphene and carbon nanostructures \cite{1} with
macroscopic bodies and microparticles is significant for the
understanding of layered systems, bundles of nanotubes or
metallic nanowires and absorption phenomena. Special
attention was attracted to this problem after the proposal of
Ref.~\cite{2} to use the single-wall carbon nanotubes for the
purposes of hydrogen storage. Subsequently  controversial
results on this subject were obtained \cite{3}. Especially, the
microscopic mechanisms underlying the absorption phenomenon remain
unclear.

Theoretically the interaction of hydrogen atoms with graphite sheets and
carbon nanotubes has been studied in Ref.~\cite{4} using the density
functional theory and in Refs.~\cite{5,6} using the nonrelativistic
perturbation theory for degenerate levels of a two-level atomic
system. The adsorption
of fullerene molecules on graphite was considered in Ref.~\cite{7}
demonstrating that the interaction is basically proportional to the
number of neighboring atoms.
 Most of theoretical work on the van der Waals force in layered
structures and between carbon nanotubes was done using the density
functional theory \cite{8,9,10,11,12,13}.
It is known, however, that in some cases density functional theory
(especially when used with linear-density approximation) gives not
enough precise description of van der Waals interactions \cite{13a}.
In Ref.~\cite{14} the
Lifshitz theory of the van der Waals and Casimir force was extended
to the case of a microparticle interacting with a plane surface
of an uniaxial crystal and with a multi-wall carbon nanotube. It was shown
that the position of a hydrogen atom inside the multi-wall carbon
nanotube is energetically preferable as compared to the outside position.

It is common knowledge that the Lifshitz theory \cite{15,16} provides
the fundamental description of the van der Waals and Casimir
interaction between two macroscopic bodies and between a microparticle
and a macroscopic body. In the framework of this theory the
interaction energy and the force are expressed in terms of the
dielectric permittivity of macrobodies or the dynamic polarizability
of a microparticle. Recently, the Lifshitz theory was successfully
applied to the interpretation of precision measurements of the Casimir
force \cite{17,18,19,20,21} and to the calculation of atom-wall
interaction in connection with Bose-Einstein condensation \cite{22,23}.
In Ref.~\cite{14} the Lifshitz theory was adapted for the case of
sufficiently thick multi-wall carbon nanotubes
using the idealized description of the wall
material by dielectric permittivity.
 This idealization is, however, not applicable for the
description of single-wall nanotubes which narrows the applicability
of the standard Lifshitz theory.

In the present paper we use the description of graphene in terms of the
two dimensional free electron gas \cite{24} in order to extend the
Lifshitz theory of the van der Waals and Casimir interaction to the
case of carbon systems. Graphene is considered as an infinitesimally
thin positively charged flat sheet, carrying a continuous fluid
with some mass and negative charge densities. This sheet is
characterized by some typical wave number $\Omega$ determined by the
parameters of the hexagonal structure of graphite. In Refs.~\cite{25,26}
the interaction of the electromagnetic oscillations with such sheet was
considered and the normal modes and reflection coefficients were found.
The van der Waals and Casimir interaction between the two parallel
plasma sheets was described in Ref.~\cite{27} using a Lifshitz-type
formula. The important distinctive feature of this formula is that
it does not use the concept of the dielectric permittivity of
the sheet.

Here we obtain Lifshitz-type formulas for the van der Waals and
Casimir interaction of graphene and material semispace or a plate.
In so doing the semispace or the plate are described using the
dielectric permittivity and graphene is considered as a plasma sheet.
The numerical computations of the interaction energy and the force
acting between graphene and thick plates made of different materials
are performed. The Lifshitz-type formula for the interaction of
graphene with a microparticle (an atom or a molecule) is also
obtained. This formula is applied to the computations of the
interaction of graphene with hydrogen atoms and molecules.
The approximate formulas for the interaction of a single-wall
carbon nanotube with a material plate are presented when the
nanotube is in close proximity or far away from the plate.

The paper is organized as follows. In Sec.~II the Lifshitz-type
formulas are obtained for the interaction of  graphene with a
semispace or a plate. Sec.~III contains the results of numerical
computations using these formulas for metal and dielectric plates.
In Sec.~IV the interaction of graphene with a hydrogen atom or a
molecule is considered.
Sec.~V  is devoted to the interaction of a single-wall carbon
nanotube with a metal or dielectric plate. Sec.~VI contains our
conclusions and discussion.

\section{Interaction of a graphene with a semispace or a plate}
We consider the van der Waals and Casimir interaction of a graphene
occupying the $xy$-plane, $z=0$, with a material semispace or a plate
of thickness $d$. The separation distance between the boundary plane
of the semispace (plate) and graphene is $a$. The dispersion
interaction of the two plane parallel bodies (plates or semispaces)
labeled by the upper indices 1 and 2 with the electromagnetic
oscillations can be described in terms of the reflection
coefficients $r_{\rm TM,TE}^{(1)}$ and $r_{\rm TM,TE}^{(2)}$ for
two independent polarizations of electromagnetic field (transverse
magnetic and transverse electric). In so doing the van der Waals
and Casimir interaction energy per unit area
is given by the Lifshitz formula
\cite{15,16,17}
\begin{eqnarray}
&&
E(a)=\frac{\hbar}{4\pi^2}\int_0^{\infty}k_{\bot}dk_{\bot}
\int_0^{\infty}d\xi\left[
\ln\left(1-r_{\rm TM}^{(1)}r_{\rm TM}^{(2)}e^{-2aq}\right)
\right.
\nonumber \\
&&\phantom{aaaaaaaaaaa}
\left.+\ln\left(1-r_{\rm TE}^{(1)}r_{\rm TE}^{(2)}e^{-2aq}
\right)\right].
\label{eq1}
\end{eqnarray}
\noindent
Here $k_{\bot}$ is the magnitude of the wave vector component
perpendicular to the $z$-axis (i.e., lying in the plane of the
plates), $\xi$ is the frequency variable along the imaginary
axis ($\omega=i\xi$), and
\begin{equation}
q=\sqrt{k_{\bot}^2+\frac{\xi^2}{c^2}}.
\label{eq2}
\end{equation}
\noindent
Equation (\ref{eq1}) is applicable at not very high temperatures
(not too large separations). As an example, at room temperature
($T=300\,$K) thermal corrections to Eq.~(\ref{eq1}) are
negligible up to separations of about 1\,$\mu$m.
If the temperature is higher (separation is larger)
Eq.~(\ref{eq1}) is simply generalized by changing the integration
with respect to $\xi$ for the summation over the discrete Matsubara
frequencies. The general derivation of Eq.~(\ref{eq1}) for
arbitrary reflection coefficients can be found in
Ref.~\cite{17}.

The commonly accepted approach (see, e.g.,
\cite{15,16,17,18,19,20,21,22,23})
uses the Fresnel reflection coefficients of the semispaces and
plates expressed in terms of the frequency-dependent dielectric
permittivities. Let the semispace labeled 2 be made of isotropic
material and be described by the dielectric permittivity
$\ee(\omega)$. In this case the reflection coefficients are
\cite{15,16,17,18,19,20,21}
\begin{eqnarray}
&&
r_{\rm TM}^{(2)}\equiv r_{\rm TM,s}^{(2)}(\xi,k_{\bot})=
\frac{\ee(i\xi)q-k}{\ee(i\xi)q+k},
\nonumber \\
&&
r_{\rm TE}^{(2)}\equiv r_{\rm TE,s}^{(2)}(\xi,k_{\bot})=
\frac{k-q}{k+q},
\label{eq3}
\end{eqnarray}
\noindent
where
\[
k=\sqrt{k_{\bot}^2+\ee(i\xi)\frac{\xi^2}{c^2}}.
\]

If the second body is a plate of finite thickness $d$,
Eq.~(\ref{eq3}) should be replaced by \cite{14}
\begin{eqnarray}
&&
r_{\rm TM}^{(2)}\equiv r_{\rm TM,p}^{(2)}(\xi,k_{\bot})=
\frac{\ee^2(i\xi)q^2-k^2}{\ee^2(i\xi)q^2+k^2+
2qk\ee(i\xi)\coth(kd)},
\nonumber \\
&&
r_{\rm TE}^{(2)}\equiv r_{\rm TE,p}^{(2)}(\xi,k_{\bot})=
\frac{k^2-q^2}{k^2+q^2+2qk\coth(kd)}.
\label{eq4}
\end{eqnarray}
\noindent
If the first body is also a semispace or a plate, its
 reflection coefficients are obtained from Eqs.~(\ref{eq3})
and (\ref{eq4}) by the replacement of the upper indices 2 for 1.

In our case the first body is graphene and it cannot be described
macroscopically it terms of the dielectric permittivity. The
van der Waals and Casimir interaction of a semispace or a plate
made of isotropic material with the graphite plate containing
a few hexagonal layers was considered in Ref.~\cite{14} using
the reflection coefficients for a uniaxial crystal. In the case
of a graphite plate of thickness $d$ they take the form
\begin{eqnarray}
&&
r_{\rm TM}^{(1)}\equiv r_{\rm TM,p}^{(1)}(\xi,k_{\bot})
\nonumber \\
&&=
\frac{\ee_x(i\xi)\ee_z(i\xi)q^2-k_z^2}{\ee_x(i\xi)\ee_z(i\xi)q^2+
k_z^2+2qk_z\sqrt{\ee_x(i\xi)\ee_z(i\xi)}\coth(k_zd)},
\nonumber \\
&&
r_{\rm TE}^{(1)}\equiv r_{\rm TE,p}^{(1)}(\xi,k_{\bot})=
\frac{k_x^2-q^2}{k_x^2+q^2+2qk_x\coth(k_xd)},
\label{eq5}
\end{eqnarray}
\noindent
where $\ee_x(\omega)=\ee_y(\omega)$ and $\ee_z(\omega)$
are the graphite dielectric
permittivities in the directions $x,y$ and $z$, respectively, and
\begin{equation}
k_x=\sqrt{k_{\bot}^2+\ee_x(i\xi)\frac{\xi^2}{c^2}},
\quad
k_z=\sqrt{k_{\bot}^2+\ee_z(i\xi)\frac{\xi^2}{c^2}}.
\label{eq6}
\end{equation}
\noindent
The number of layers in the graphite plate should be sufficiently large
that the macroscopic description in terms of the dielectric permittivities
is applicable.

The reflection coefficients for a sheet of graphene  cannot be obtained
from Eq.~(\ref{eq5}) in the limit $d\to 0$ [in fact, the coefficients
(\ref{eq5}) go to zero when $d$ vanishes]. The reason is that the
case of ``thin'' plate implies that $d/a$ is sufficiently small,
whereas $d$ should be large enough for the validity of the
macroscopic description in terms of $\ee$. In Sec.~III the correlation
between the macroscopic description of a graphite plate by means
of Eq.~(\ref{eq5}) and the case of graphene will be clarified.

A single plane hexagonal layer of graphite, i.e., graphene, can
be described as infinitely thin plasma sheet where the $\pi$-electrons are
treated as a continuously charged fluid moving in an immobile, overall
neutralizing background of  positive charge. These plasma sheets have
been considered in Ref.~\cite{24} and, more recently, by Barton
\cite{25,26} in connection with the Casimir effect for a fullerene and
for a single base plane from graphite. In Ref.~\cite{27} they were
used to calculate the van der Waals and Casimir interaction between the
two parallel graphenes and a Lifshitz-type formula was obtained for
their interaction energy.

The plasma sheet model assumes the $\pi$-electrons of the carbon
atoms to be described by a negatively charged fluid confined to a
plane and having the two dimensional displacement vector
$\mbox{\boldmath$R$}(x,y)\exp(-i\omega t)$, where we introduced
at once the usual harmonic time dependence.
(In \cite{25,26} this vector is denoted by $\xi$ and the wave number
of the sheet $\Omega$ by $q$.)
 The fluid has a surface
charge density $ne$ and a surface mass density $nm$ where $e$ and $m$
are the electron charge and mass, respectively. For the hexagonal
structure of carbon layers there is one $\pi$-electron per
atom \cite{28} resulting in two $\pi$-electrons per one hexagonal cell.
This leads to
\begin{equation}
n=\frac{4}{3\sqrt{3}l^2},
\label{eq7}
\end{equation}
\noindent
where $l=1.421\,${\AA} is the side length of a hexagon.

The fluid provides a source for the Maxwell equations with  surface
charge and surface current densities,
\begin{equation}
\sigma=-ne\nabla_{\!t}\cdot\mbox{\boldmath$R$},
\quad
\mbox{\boldmath$j$}=-i\omega ne\mbox{\boldmath$R$},
\label{eq8}
\end{equation}
\noindent
where the operator $\nabla_{\!t}$ acts in the tangential direction
to the sheet [here and below $\sigma,\,\mbox{\boldmath$j$}$ and
fields $\mbox{\boldmath$E$}$ and $\mbox{\boldmath$B$}$ depend on
coordinates; their dependence on time is obtained through the
multiplication by the common factor $\exp(-i\omega t)$].
The Maxwell equations read
\begin{eqnarray}
&&
\nabla\cdot\mbox{\boldmath$E$}=4\pi\sigma\delta(z),\quad
\nabla\times\mbox{\boldmath$E$}-\frac{i\omega}{c}\mbox{\boldmath$B$}=0,
\label{eq9} \\
&&
\nabla\cdot\mbox{\boldmath$B$}=0,\quad
\nabla\times\mbox{\boldmath$B$}+\frac{i\omega}{c}\mbox{\boldmath$E$}=
\frac{4\pi}{c}\mbox{\boldmath$j$}\delta(z).
\nonumber
\end{eqnarray}
\noindent
By the integration of Maxwell equations across the sheet, we
obtain the matching conditions on the tangential and normal
components of the fields \cite{26},
\begin{eqnarray}
&&
\mbox{\boldmath$E$}_{t,2}-\mbox{\boldmath$E$}_{t,1}=0, \quad
E_{z,2}-E_{z,1}=2\Omega\frac{c^2}{\omega^2}
\nabla_{\! t}\cdot\mbox{\boldmath$E$}_t,
\label{eq10} \\
&&
B_{z,2}-B_{z,1}=0, \quad
\mbox{\boldmath$B$}_{t,2}-\mbox{\boldmath$B$}_{t,1}=
-2i\Omega\frac{c}{\omega}\mbox{\boldmath$z$}
\times\mbox{\boldmath$E$}_{t}.
\nonumber
\end{eqnarray}
\noindent
Here $\mbox{\boldmath$z$}=(0,\,0,\,1)$ is the unit vector pointing
in $z$-direction,
\begin{equation}
\Omega=2\pi\frac{ne^2}{mc^2}=6.75\times10^5\,\mbox{m}^{-1}
\label{eq11}
\end{equation}
\noindent
and $n$ is defined in Eq.~(\ref{eq7}). The quantity
$\Omega$ is the main
characteristic of graphene in the model under consideration.
The value in Eq.~(\ref{eq11}) corresponds to the frequency
$\omega_{\Omega}=c\Omega=2.02\times 10^{14}\,$rad/s
(as compared, for instance, to the plasma frequency of gold
$\omega_p=1.37\times 10^{16}\,$rad/s).

We remark that now the matching conditions (\ref{eq10}) together
with the Maxwell equations (\ref{eq9}) outside the surface, i.e.,
without the delta functions on the right-hand sides, provide
the complete description of the interaction of the electromagnetic
field with the plasma sheet. This implies, for instance, that
all components of the field satisfy the usual Poisson equations,
\begin{equation}
\left(\Delta+\frac{\omega^2}{c^2}\right)\mbox{\boldmath$E$}=0,
\quad
\left(\Delta+\frac{\omega^2}{c^2}\right)\mbox{\boldmath$B$}=0.
\label{eq12}
\end{equation}

After the separation of variables $x$ and $y$ in Eqs.~(\ref{eq10})
and (\ref{eq12}) we arrive at a one-dimensional scattering problem
in the $z$-direction \cite{17}. The solution of this problem leads to
the following reflection coefficients on the graphene plasma sheet
taken at the imaginary frequency axis \cite{26},
\begin{eqnarray}
&&
r_{\rm TM}^{(1)}\equiv r_{\rm TM,g}^{(1)}(\xi,k_{\bot})=
\frac{c^2q\Omega}{c^2q\Omega+\xi^2},
\nonumber \\
&&
r_{\rm TE}^{(1)}\equiv r_{\rm TE,g}^{(1)}(\xi,k_{\bot})=
\frac{\Omega}{\Omega+q}.
\label{eq13}
\end{eqnarray}

Equations (\ref{eq1}) and (\ref{eq3}), (\ref{eq13}) or,
alternatively, (\ref{eq1}) and (\ref{eq4}), (\ref{eq13}) allow
to calculate the energy of the van der Waals and Casimir
interaction between a graphene and a semispace or a plate made
of some usual material described by the dielectric permittivity
$\ee(\omega)$. From Eq.~(\ref{eq1}) it is easy to obtain the
Lifshitz-type formula for the van der Waals and Casimir force
per unit area
acting between graphene and material semispace or a plate,
\begin{eqnarray}
&&
F(a)=-\frac{\partial E(a)}{\partial a}=
-\frac{\hbar}{2\pi^2}\int_0^{\infty}qk_{\bot}dk_{\bot}
\int_0^{\infty}d\xi
\nonumber \\
&&\phantom{aaa}
\times\left(\frac{r_{\rm TM}^{(1)}r_{\rm TM}^{(2)}}{e^{2aq}-
r_{\rm TM}^{(1)}r_{\rm TM}^{(2)}}+
\frac{r_{\rm TE}^{(1)}r_{\rm TE}^{(2)}}{e^{2aq}-
r_{\rm TE}^{(1)}r_{\rm TE}^{(2)}}
\right).
\label{eq14}
\end{eqnarray}

Equations (\ref{eq1}) and (\ref{eq14}) can be used to obtain
the interaction energy and the force between an atom (molecule)
and a graphene sheet (see Sec.~IV) and between a carbon nanotube
and a material wall (Sec.~V). In the next section these equations
are applied to compute the interaction of graphene with metal and
dielectric walls made of real materials.

\section{Computations of the van der Waals and Casimir
interaction between graphene and metal or dielectric plate}
Here we apply Eqs.~(\ref{eq1}) and (\ref{eq14}) to calculate
the van der Waals and Casimir interaction energy and force acting
between graphene and thick plates (semispaces) made of different
materials. The reflection coefficients $r_{\rm TM,TE}^{(1)}$
for graphene are given in Eq.~(\ref{eq13}), and the
reflection coefficients $r_{\rm TM,TE}^{(2)}$ for a semispace
are presented in Eq.~(\ref{eq3}). All parameters in Eq.~(\ref{eq13})
are specified. As to Eq.~(\ref{eq3}), it depends on the dielectric
permittivity along the imaginary frequency axis. In our
computations we consider metal and dielectric semispaces made of
Au and high resistivity Si, respectively. High precision results
for the dielectric permittivities of these materials along the
imaginary frequency axis were obtained in Ref.~\cite{29} by
means of the Kramers-Kronig relation
\begin{equation}
\ee(i\xi)=1+\frac{2}{\pi}\int_{0}^{\infty}d\omega
\frac{\omega\mbox{Im}\ee(\omega)}{\omega^2+\xi^2}.
\label{eq15}
\end{equation}
\noindent
The imaginary part of the dielectric permittivities at real
frequencies was taken from the tabulated optical data \cite{30}.

The computational results for the van der Waals and Casimir
energy density $E(a)$ normalized to the Casimir energy density
in the configuration of the ideal metal plates,
\begin{equation}
E_0(a)=-\frac{\pi^2}{720}\,\frac{\hbar c}{a^3},
\label{eq16}
\end{equation}
\noindent
are shown in Fig.~1(a) as a function of separation.
The solid and dashed lines are related to the interaction of
graphene with Au and Si, respectively. In Fig.~1(b) the
analogous results for the van der Waals and Casimir force
per unit area $F(a)$ normalized to the force per unit area,
\begin{equation}
F_0(a)=-\frac{\pi^2}{240}\,\frac{\hbar c}{a^4},
\label{eq17}
\end{equation}
\noindent
acting between ideal metals are presented.

As is seen in Fig.~1, both the relative interaction energy and
force between graphene and Au are greater than between graphene
and Si at all separations. With the decrease of separation these
quantities decrease. This is because $E(a)$ and $F(a)$ go to
infinity more slowly than the respective dependencies for ideal
metals in Eqs.~(\ref{eq16}) and (\ref{eq17}). As a result, at the
shortest separation indicated in Fig.~1 ($a=3\,$nm)
$E/E_0=0.0252$, $F/F_0=0.0203$ for Au and
$E/E_0=0.0212$, $F/F_0=0.0170$ for Si.
Note that at separations less than 1--2\,nm the
used model of graphene as a plasma sheet may become inapplicable
because it does not take the atomic structure into account.
With the increase of separation up to 1\,$\mu$m the interaction
energy between graphene and Au achieves almost one half of that
between ideal metals and the force magnitude is larger than 0.43
of that in the case of ideal metals.

Now we compare  the above numerical results for Au with the
analytic calculations in the asymptotic region of short
separations. The dielectric permittivity of Au along the imaginary
frequency axis can be approximated by means of the plasma model,
\begin{equation}
\ee(i\xi)=1+\frac{\omega_p^2}{\xi^2},
\label{eq18}
\end{equation}
\noindent
where $\omega_p$ is the plasma frequency. At short separations only
the TM mode in Eq.~(\ref{eq1}) contributes essentially to the
van der Waals energy \cite{16}. Introducing the dimensionless
variables $v=aq$ and $\zeta=a\xi/c$ and substituting Eq.~(\ref{eq18})
in Eq.~(\ref{eq3}), we rewrite the TM contribution to Eq.~(\ref{eq1})
in the following way:
\begin{eqnarray}
&&
E(a)=E_0(a)f(\alpha,\beta),
\label{eq19} \\
&&
f(\alpha,\beta)=-\frac{180}{\pi^4}\int_0^{\infty}vdv\int_0^v
d\zeta
\nonumber \\
&&\phantom{aaaaaa}
\times\ln\left[1-\frac{\alpha v}{\alpha v+\zeta^2}\,
\frac{\left(1+
\frac{\beta^2}{\zeta^2}\right)v-\sqrt{\beta^2+v^2}}{\left(1+
\frac{\beta^2}{\zeta^2}\right)v+\sqrt{\beta^2+v^2}}\,
e^{-2v}\right].
\nonumber
\end{eqnarray}
\noindent
Here, $E_0(a)$ was defined in Eq.~(\ref{eq16}) and the two parameters are
$\alpha=\Omega a$ and $\beta=\omega_pa/c$. Instead of the variable
$\zeta$ we introduce the variable $\eta=\zeta/(\beta v)$. This
brings the correction factor $f$ in Eq.~(\ref{eq19}) to the form

\begin{eqnarray}
&&
f(\alpha,\beta)=-\frac{180\beta}{\pi^4}\int_0^{\infty}v^2dv\int_0^{1/\beta}
d\eta
\label{eq20} \\
&&\phantom{aaaaaa}
\times\ln\left[1-\frac{e^{-2v}}{1+\frac{v\eta^2}{t}}\,
\frac{\left(1+
\frac{1}{v^2\eta^2}\right)v-\sqrt{\beta^2+v^2}}{\left(1+
\frac{1}{v^2\eta^2}\right)v+\sqrt{\beta^2+v^2}}\right],
\nonumber
\end{eqnarray}
\noindent
where $t\equiv\alpha/\beta^2$. At short separations the parameters
$\alpha$ and $\beta$ are small (at the shortest separation $a=3\,$nm
it holds $\alpha=2.02\times 10^{-3}$, $\beta=0.137$).
It is easily seen that only small values of $\eta$ give the major contribution
to the integral with respect to $\eta$ in Eq.~(\ref{eq20}). Because of this,
without loss of accuracy, one can replace the upper limit $1/\beta$
for $\infty$. Expanding the integrand in  Eq.~(\ref{eq20}) in powers of
$\beta$, one arrives at
\begin{equation}
f(\alpha,\beta)=\beta h_0(t)+\beta^3h_1(t)+\ldots\, ,
\label{eq21}
\end{equation}
\noindent
where
\begin{eqnarray}
&&
h_0(t)=-\frac{180}{\pi^4}\int_0^{\infty}v^2dv\int_0^{\infty}
d\eta
\label{eq22} \\
&&\phantom{aaaaaa}
\times\ln\left[1-\frac{e^{-2v}}{\left(1+
\frac{v\eta^2}{t}\right)\left(1+2v^2\eta^2\right)}\right].
\nonumber
\end{eqnarray}
\noindent
Notice that the term of order $\beta^3$ on the right-hand side
of Eq.~(\ref{eq21}) and the respective terms originating from
the TE contribution to Eq.~(\ref{eq1}) are small at short
separations and can be neglected. The parameter $t=\alpha/\beta^2$
is small and decreases with the increase of separation starting
from the largest value of $t=0.108$ at the shortest separation
$a=3\,$nm. In Fig.~2 the values of $h_0$ are plotted as a function
of $t$.

As a result, the asymptotic representation of the van der Waals
interaction energy between a graphene and an Au semispace at
short separations is given by
\begin{equation}
E(a)=E_0(a)\frac{\omega_pa}{c}\,
h_0\left(\frac{\Omega c^2}{\omega_p^2a}\right).
\label{eq23}
\end{equation}
\noindent
The comparison of Eq.~(\ref{eq23}) with the results of numerical
computations in Fig.~1(a) (solid line) shows good agreement
in the limits of 3\% within the separation region from 15 to
80\,nm. At the shortest separation of 3\,nm the error of the
representation (\ref{eq23}) achieves 11.5\% if to compare with the
computational results using the tabulated optical data.
At the same time, within the separation region from 3 to 40\,nm,
Eq.~(\ref{eq23}) is in agreement with the
numerical computations using the plasma model (\ref{eq18}) up to a
maximal error of 3\%.
A few computational results are presented in Table~1.
Column 1 contains separation distances, and in columns 2, 3 and 4
the values of the correction factor $E(a)/E_0(a)$ are contained
computed by using the tabulated optical data, the plasma model
dielectric function and the asymptotic representation (\ref{eq23}),
respectively.

Note that the  asymptotic expression (\ref{eq23}) at
short separations contains the velocity of light, i.e., is
relativistic. This is different from the van der Waals
interaction at short separations of two bodies described by the
dielectric permittivity \cite{16} but is in accordance with
the short separation interaction of the two plasma sheets considered
in Ref.~\cite{27}.

It is instructive to consider the relationship between the graphene
described by the model of a plasma sheet and the thin graphite plate
containing a few hexagonal layers. For this purpose we compare the
computed above interaction energy of graphene and Au semispace with
the interaction of a graphite plate of some thickness $d$ and the
same semispace. The latter is computed using Eq.~(\ref{eq1}) with
the reflection coefficients (\ref{eq3}) and (\ref{eq5}).
The dielectric permittivities of graphite along the imaginary
frequency axis $\ee_{x,y}(i\xi)$ and $\ee_{z}(i\xi)$ were computed
in Ref.~\cite{14} by means of the Kramers-Kronig relations using
the tabulated optical data for graphite from Ref.~\cite{30}.
Here we use the results of that computations.

In Fig.~3 the interaction energy density of the graphite plate
and Au semispace normalized for the case of ideal metals in
Eq.~(\ref{eq16}) is plotted as a function of the relative plate
thickness $d/a$ at separations $a=1,\,0.5,\,0.3,\, 0.1,$ and
0.05\,$\mu$m (lines 1,\,2,\,3,\,4, and 5, respectively).
As is seen in Fig.~3, the magnitudes of $E/E_0$ decrease
when $d/a\to 0$. This is in accordance with Eq.~(\ref{eq5}) for
the reflection coefficients which go to zero when $d$ vanishes.
{}From this it follows that the graphene cannot be obtained from the
macroscopic plate in the limit of zero thickness. There is, however,
some relationship between the cases of graphene and the graphite
plate. To trace it let us plot the values of graphene-Au relative
interaction energy at separations $1,\,0.5,\,0.3,\, 0.1,$ and
0.05\,$\mu$m from Fig.~1(a) (the solid line) on the vertical axis
in Fig.~3 and draw the tangents to the respective lines
 1,\,2,\,3,\,4, and 5. It is seen that all tangents intersect with
lines $1,\,\ldots\,,\,5$ in the region $d/a\sim 0.13-0.18$.
These values of the relative thickness are in fact characteristic
for the graphite plates could be treated macroscopically in terms
of the dielectric permittivity. As an example, at a separation of
50\,nm (line 5 in Fig.~3) only the plates with thickness
$d\geq 0.15\cdot 50=7.5\,$nm are enough thick to be considered
macroscopically.

\section{Interaction of graphene with a hydrogen atom or a molecule}
In Sec.~II we have described the interaction of graphene with a
material semispace by means of the Lifshitz-type formula (\ref{eq1})
with the reflection coefficients (\ref{eq3}) and (\ref{eq13}).
This permits us to derive a Lifshitz-type formula for an atom
near graphene by considering the limit of rarefied semispace
\cite{15,16}. Let us expand the dielectric permittivity of a
semispace in powers of the number of atoms per unit volume $N$
preserving only the first order contribution \cite{15},
\begin{equation}
\ee(i\xi)=1+4\pi\alpha(i\xi)N+\mbox{O}(N^2),
\label{eq41}
\end{equation}
\noindent
where $\alpha(\omega)$ is the dynamic polarizability of an atom.

Expanding the reflection coefficients (\ref{eq3}) and the energy
density (\ref{eq1}) in powers of $N$ (see Ref.~\cite{22} for
details), we arrive at the energy of the interaction between
an atom and a graphene,
\begin{eqnarray}
&&
E^{A}(a)=-\frac{\hbar}{2\pi}\int_0^{\infty}k_{\bot}dk_{\bot}
\int_0^{\infty}d\xi\alpha(i\xi)qe^{-2aq}
\label{eq42} \\
&&\phantom{aaaaaa}
\times\left[2r_{\rm TM,g}^{(1)}+\frac{\xi^2}{q^2c^2}
\left(r_{\rm TE,g}^{(1)}-r_{\rm TM,g}^{(1)}\right)\right].
\nonumber
\end{eqnarray}
\noindent
Recall that the reflection coefficients of graphene are given
by Eqs.~(\ref{eq13}).

As was shown in Ref.~\cite{14}, the atomic and molecular dynamic
polarizabilities of H can be represented with sufficient precision
using the single oscillator model,
\begin{eqnarray}
&&
\alpha(i\xi)=\alpha_a(i\xi)=\frac{g_a}{\omega_a^2+\xi^2},
\label{eq43} \\
&&
\alpha(i\xi)=\alpha_m(i\xi)=\frac{g_m}{\omega_m^2+\xi^2}.
\nonumber
\end{eqnarray}
\noindent
Here, $g_a\equiv\alpha_a(0)\omega_a^2$ is expressed through the
static atomic polarizability $\alpha_a(0)=4.50\,$a.u. and the
characteristic frequency $\omega_a=11.65\,$eV \cite{31}.
For a hydrogen molecule  $g_m\equiv\alpha_m(0)\omega_m^2$
where $\alpha_m(0)=5.439\,$a.u. and $\omega_m=14.09\,$eV \cite{31}.
Note that before the substitution in Eq.~(\ref{eq42}) the
polarizabilities (\ref{eq43}) should be expressed in cubic meters
taking into account that
$\mbox{1\,a.u. of polarizability}=1.482\times 10^{-31}\,\mbox{m}^3$.

For convenience in numerical computations, we introduce the dimensionless
variable $y=2aq$ and represent Eq.~(\ref{eq42}) in terms of the
van der Waals coefficient $C_3$:
\begin{eqnarray}
&&
E^{A}(a)=-\frac{C_3(a)}{a^3},
\label{eq44} \\
&&
C_3(a)=\frac{\hbar}{16\pi}\int_0^{\infty}dye^{-y}
\int_0^{cy/(2a)}d\xi\alpha(i\xi)
\nonumber \\
&&\phantom{aaaaaa}
\times\left[2y^2r_{\rm TM,g}^{(1)}+\frac{4a^2\xi^2}{c^2}
\left(r_{\rm TE,g}^{(1)}-r_{\rm TM,g}^{(1)}\right)\right].
\nonumber
\end{eqnarray}

The computational results for $C_3$ as a function of $a$ obtained by
using Eqs.~(\ref{eq43}) and (\ref{eq44}) are presented in Fig.~4.
The solid and dashed lines are related to the interaction of
graphene with hydrogen atom and molecule, respectively.
As is seen in Fig.~4, the magnitudes of the van der Waals coefficient
for the hydrogen molecule interacting with graphene are larger than
for the atom at all separations. Note that
$\mbox{1\,a.u.{\ }of{\ }}C_3=0.646\times 10^{-48}\,\mbox{J\,m}^3$.
The comparison with the results of Ref.~\cite{14} for the interaction
of H atoms and molecules with graphite semispace shows that the
interaction with graphene is by a factor of more than 2.5 weaker.

\section{Interaction of single-wall carbon nanotubes with
metal or dielectric plate}
Let the single-wall nanotube of radius $R$ lie along
the $y$ axis at a separation $a$ from the boundary surface of
a material semispace. For sufficiently small $a$ the interaction
energy in such configuration can be approximately obtained by
using the proximity force theorem \cite{32} and the Lifshitz-type
formula (\ref{eq1}) for the van der Waals energy between a graphene
and a material semispace. According to the proximity force theorem,
we replace the cylindrical surface by a set of infinitely
long plane strips of width $dx$. The interaction between each strip,
substituting a part of the cylindrical surface, and the opposite
strip belonging to the boundary plane of a semispace is calculated
by using Eq.~(\ref{eq1}). The separation distance between the two
opposite strips with coordinate $x$ is
\begin{equation}
z=z(x)=a+R-\sqrt{R^2-x^2}.
\label{eq51}
\end{equation}
\noindent
Expanding the logarithms in Eq.~(\ref{eq1}) in a power series, we present
the interaction energy density between the strips in the form
\begin{eqnarray}
&&
E[z(x)]=-\frac{\hbar}{4\pi^2}\int_0^{\infty}k_{\bot}dk_{\bot}
\int_0^{\infty}d\xi\sum\limits_{n=1}^{\infty}\frac{1}{n}
\label{eq52}\\
&&\phantom{aaaaaaaaaa}
\times\left[\left(r_{\rm TM}^{(1)}r_{\rm TM}^{(2)}\right)^n+
\left(r_{\rm TE}^{(1)}r_{\rm TE}^{(2)}\right)^n\right]\,
e^{-2z(x)qn}.
\nonumber
\end{eqnarray}
\noindent
Recall that the reflection coefficients in the case under consideration
are given by Eqs.~(\ref{eq3}) and (\ref{eq13}).

To find the interaction energy $E^n$ per unit length between the
semispace and the nanotube, we integrate Eq.~(\ref{eq52}) from
$x=-R$ to $x=R$ (this is equal to twice the integral from zero to $R$).
In so doing we replace the variable $k_{\bot}$ with $q$ from
Eq.~(\ref{eq2}):
\begin{eqnarray}
&&
E^n(a)=-\frac{\hbar}{2\pi^2}\int_0^{\infty}qdq
\int_0^{cq}d\xi\sum\limits_{n=1}^{\infty}\frac{1}{n}
e^{-2aqn}
\label{eq53}\\
&&
\times\left[\left(r_{\rm TM}^{(1)}r_{\rm TM}^{(2)}\right)^n+
\left(r_{\rm TE}^{(1)}r_{\rm TE}^{(2)}\right)^n\right]\,
\int_{0}^{R}dxe^{-2qn(R-\sqrt{R^2-x^2})}.
\nonumber
\end{eqnarray}
\noindent
By introducing the new variable $s=1-\sqrt{1-x^2/R^2}$, the
integral $I$ with respect to $x$ in Eq.~(\ref{eq53})
can be written in the form
\begin{equation}
I=R\int_{0}^{1}\frac{(1-s)ds}{\sqrt{s(2-s)}}
e^{-2qnRs}.
\label{eq54}
\end{equation}
\noindent
The major contribution in Eqs.~(\ref{eq53}) and (\ref{eq54})
comes from $q\sim 1/a$. Bearing in mind that the proximity force
theorem works good for $R\gg a$, we conclude that the magnitude
of the integral $I$ is determined by the behavior of the
integrand around the lower integration limit. Neglecting $s$
compared with unity in Eq.~(\ref{eq54}), we arrive at
\begin{equation}
I=\frac{R}{\sqrt{2}}\int_{0}^{1}\frac{e^{-2qnRs}}{\sqrt{s}}ds
=\frac{1}{2}\sqrt{\frac{\pi R}{qn}}\mbox{Erf}(\sqrt{2qnR}),
\label{55}
\end{equation}
\noindent
where Erf$(z)$ is the error function. Using once more the
conditions $q\sim 1/a$ and $R\gg a$, we conclude that
$\mbox{Erf}(\sqrt{2qnR})\approx 1$ and obtain from  Eq.~(\ref{eq53})
\begin{eqnarray}
&&
E^n(a)=-\frac{\hbar\sqrt{\pi R}}{4\pi^2}\int_0^{\infty}\sqrt{q}dq
\int_0^{cq}d\xi\sum\limits_{n=1}^{\infty}\frac{1}{n\sqrt{n}}
e^{-2aqn}
\nonumber\\
&&\phantom{aaaaaaaaa}
\times\left[\left(r_{\rm TM}^{(1)}r_{\rm TM}^{(2)}\right)^n+
\left(r_{\rm TE}^{(1)}r_{\rm TE}^{(2)}\right)^n\right]
\label{eq56}\\
&&\phantom{aaaaa}
=-\frac{\hbar\sqrt{R}}{4\pi^{3/2}}\int_0^{\infty}\sqrt{q}dq
\int_0^{cq}d\xi
\nonumber\\
&&\phantom{aaaaaa}\times\left[\mbox{Li}_{3/2}
\left(r_{\rm TM}^{(1)}r_{\rm TM}^{(2)}e^{-2aq}\right)+
\mbox{Li}_{3/2}\left(r_{\rm TE}^{(1)}r_{\rm TE}^{(2)}
e^{-2aq}\right)\right],
\nonumber
\end{eqnarray}
\noindent
where $\mbox{Li}_{3/2}(z)$ is the polylogarithm function.

By analogy, for the force per unit length it follows
\begin{eqnarray}
&&F^n(a)=
-\frac{\hbar\sqrt{R}}{2\pi^{3/2}}\int_0^{\infty}{q}^{3/2}dq
\int_0^{cq}d\xi
\label{eq57}\\
&&\phantom{aaaaa}\times\left[\mbox{Li}_{1/2}
\left(r_{\rm TM}^{(1)}r_{\rm TM}^{(2)}e^{-2aq}\right)+
\mbox{Li}_{1/2}\left(r_{\rm TE}^{(1)}r_{\rm TE}^{(2)}
e^{-2aq}\right)\right].
\nonumber
\end{eqnarray}
\noindent
Note that  for an ideal metal cylinder in close proximity to
an ideal metal plate the interaction energy and force per unit
length are given by \cite{33}
\begin{equation}
E_0^n(a)=-\frac{1}{a^2}\sqrt{\frac{R}{a}}
\frac{\pi^3\hbar c}{960\sqrt{2}},
\quad
F_0^n(a)=-\frac{1}{a^3}\sqrt{\frac{R}{a}}
\frac{\pi^3\hbar c}{384\sqrt{2}}.
\label{eq58}
\end{equation}

It is significant that Eqs.~(\ref{eq56}) and (\ref{eq57}) follow also
in the limit of short separations from the exact result for a nanotube
described by the cylindrical surface with the boundary conditions
(\ref{eq10}) located above a material semispace. This result can be
obtained by the method of functional determinants used previously in
Refs.~\cite{33,34,35} for the configurations of a sphere or a cylinder
made of ideal metal placed above an ideal metal plate (the details will
be published elsewhere). The exact result contains also the
asymptotic case of large separations $a\gg R$ between the nanotube
and the metal semispace described by the plasma model (\ref{eq18}):
\begin{equation}
E^n(a)=\frac{\hbar c}{2\pi a^2\ln\frac{R}{2a}}
\int_0^{\infty}\rho d\rho
\frac{\sqrt{\left(\frac{\omega_pa}{c}\right)^2+\rho^2}-
\rho}{\sqrt{\left(\frac{\omega_pa}{c}\right)^2+\rho^2}+
\rho}\,e^{-2\rho}.
\label{eq59}
\end{equation}
\noindent
This asymptotic behavior does not depend on the nanotube parameter $\Omega$.
In the limit $\omega_p\to\infty$ the integral in Eq.~(\ref{eq59})
tends to 1/4 and we arrive at the previously known result
\begin{equation}
E_0^n(a)=\frac{\hbar c}{8\pi a^2\ln\frac{R}{2a}},
\label{eq510}
\end{equation}
\noindent
valid for an ideal metal cylinder far away from an ideal metal plate
\cite{34,33}.

Now we present the computational results for a nanotube in close
proximity and far away from a semispace. Consider, first, the
nanotube of radius $R$ in close proximity to Au or Si semispace and
compute the normalized interaction energy $E^n/E_0^n$ and the force
$F^n/F_0^n$ using Eqs.~(\ref{eq56})--(\ref{eq58}). As in Sec.~III,
the dielectric permittivities of Au and Si along the imaginary
frequency axis are found from the Kramers-Kronig relations and
tabulated optical data for the complex index of refraction.
A few computed results are presented in Table~II. Column 1
contains the separation distance, columns 2 and 3 contain the values of
the correction factor $E^n(a)/E_0^n(a)$ for Au and Si, respectively,
and columns 4 and 5 contain the analogous  correction factor
to the force $F^n(a)/F_0^n(a)$. As is seen in Table~II, the
magnitudes of all correction factors for Si are smaller than for Au
and they are monotonously increasing functions with the increase
of separation. Note that the normalized values in Table~II do not depend
on the nanotube radius. However, bearing in mind that the largest
diameter of a single-wall carbon nanotube is of about 10\,nm
\cite{1}, the separation region where the approximate equations
(\ref{eq56}) and (\ref{eq57}) are applicable is very narrow.

Now we consider the nanotube of radius $R$ far away from an Au surface
described by the plasma model ($a\gg R$). In this case the Casimir
interaction energy is given by  Eq.~(\ref{eq59}). The results of
numerical computations of the normalized interaction energy
$E^n(a)/E_0^n(a)$, where the energy $E_0^n(a)$ for an ideal metal
case is given in Eq.~(\ref{eq510}), are presented in Fig.~5 as
a function of separation. As is seen in Fig.~5, the normalized interaction
energy increases with the increase of separation approaching  the case
of an ideal metal cylinder above an ideal metal plate. At $a=1\,\mu$m
it achieves the value of 0.956 very close to the ideal metal case.
For the typical single-wall
nanotube diameter of about 1\,nm the asymptotic expression
(\ref{eq59}) is applicable in the wide separation region from a few
nanometers to about 1 micrometer (for larger separations the thermal
corrections should be taken into account).

\section{Conclusions and discussion}
In the above we have obtained Lifshitz-type formulas describing
the van der Waals and Casimir interaction between graphene and material
plate,  between an atom or a molecule and graphene, and between
a single-wall carbon nanotube and material plate. The distinguishing
feature of these formulas is that they describe graphene by using the
reflection coefficients obtained from the specific boundary conditions
for the electromagnetic oscillations on the infinitely thin plasma
sheet. This permits to circumvent the use of the concept of dielectric
permittivity commonly used in the Lifshitz theory of the van der Waals
and Casimir force between macrobodies, but being not directly applicable
to single-wall carbon nanostructures. The developed formalism is
supplementary to widely applied theoretical approaches describing the
interactions of carbon nanostructures with macrobodies, atoms or
molecules, e.g., to the density functional theory (see Introduction).
Together with Ref.~\cite{14}, where the Lifshitz theory was applied
to the multi-wall carbon nanotubes, it provides the foundation for the
application of quantum statistical physics to the investigation of
dispersion forces in carbon nanostructures.

The obtained Lifshitz-type formulas for the van der Waals and Casimir
energy and force were applied to the case of graphene interacting with
Au and Si walls. The wall material was described by the dielectric
permittivity along the imaginary frequency axis computed using the
tabulated optical data for the complex index of refraction for Au and
Si. In the case of Au the analytic asymptotic expression for the
interaction energy with graphene was also obtained using the plasma
model dielectric function. The relationship between the graphene,
described by the plasma sheet, and the thin graphite plate was
investigated. As an example of microparticle-carbon nanostructure
interaction, we have calculated the van der Waals coefficients for
the interaction of hydrogen atoms and molecules with graphene.
Another example, considered in the paper, is the van der Waals
or Casimir interaction of a material wall with a single-wall
carbon nanotube in close proximity or far away from the wall,
respectively. As was noted in Sec.~III, the obtained results for
the interaction energy and force are applicable for practical
calculations at separations larger than 1--2\,nm. At smaller
separations there may be the attractive forces of chemical nature.
At separations less than 1\,nm short-range repulsive forces of
exchange nature come into play. These forces depend on atomic
structure of a surface and cannot be described macroscopically
by means of the boundary conditions. At intermediate separations
between the exchange repulsion and van der Waals attraction some
phenomenological potentials can be used for practical calculations
\cite{36a}.

In the future work it would be interesting to perform the comparative
computations of the van der Waals interaction with carbon nanostructures
by using different theoretical approaches. The suggested application
of the Lifshitz theory to graphene and single-wall carbon nanotubes
can be extended to more complicated structures such as fullerene
molecules or graphitic cones \cite{36}. The interaction of
hydrogen atoms and
molecules with the single-wall carbon nanotubes can be also considered on
the same footings.
This can be done approximately by using the proximity force theorem
because the immediate application of the Lifshitz-type formulas is
possible for only planar structures (the results of this work in
progress will be published elsewhere).
In this paper we did not deal with the thermal
effects. However, the generalization of the proposed formalism to the
case of nonzero temperatures is straightforward.

\section*{Acknowledgments}
G.~L.~K. and V.~M.~M. are grateful to the Center of Theoretical
Studies and the Institute for Theoretical Physics, Leipzig University for
their kind hospitality. This work was supported by Deutsche
Forschungsgemeinschaft, Grant No. 436 RUS 113/789/0-2.
M.B.'s work was supported by the research funding from
the EC's Sixth Framework Programme within the STRP
project ``PARNASS" (NMP4-CT-2005-01707).
G.~L.~K. and V.~M.~M. were also partially supported by
the Russian Foundation for
Basic Research (Grant No. 05--08--18119a).

\begin{figure*}
\vspace*{-2cm}
\includegraphics{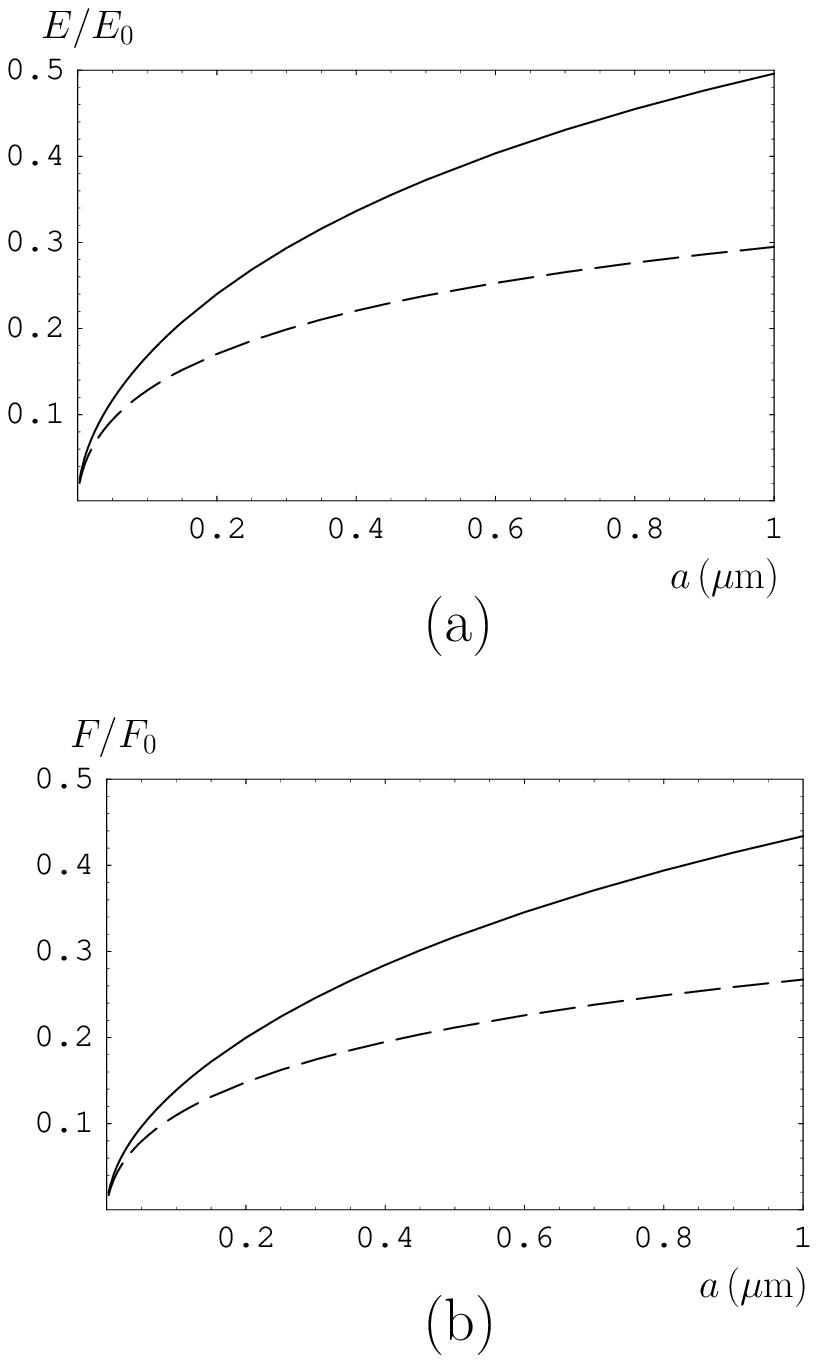}
\vspace*{-11cm}
\caption{The normalized to the case of ideal metals van der Waals
and Casimir energy (a) and force (b) per unit area between a graphene
and a semispace versus separation. The solid and dashed lines are
related to the semispace made of Au and Si, respectively.
}
\end{figure*}
\begin{figure*}
\vspace*{-7cm}
\includegraphics{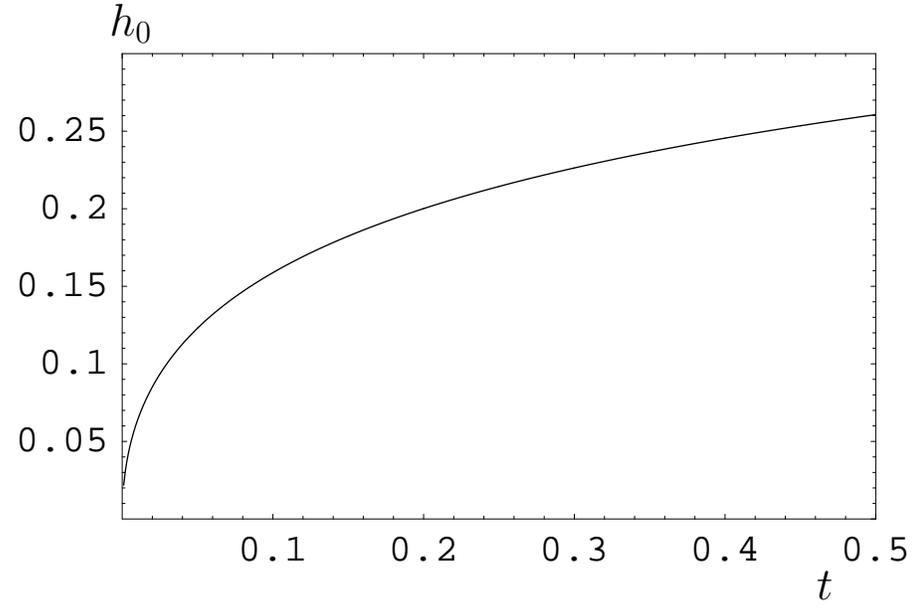}
\vspace*{-10cm}
\caption{The function $h_0$ in Eq.~(\ref{eq21}) versus the
dimensionless parameter $t=\Omega c^2/(\omega_p^2a)$.
}
\end{figure*}
\begin{figure*}
\vspace*{-7cm}
\includegraphics{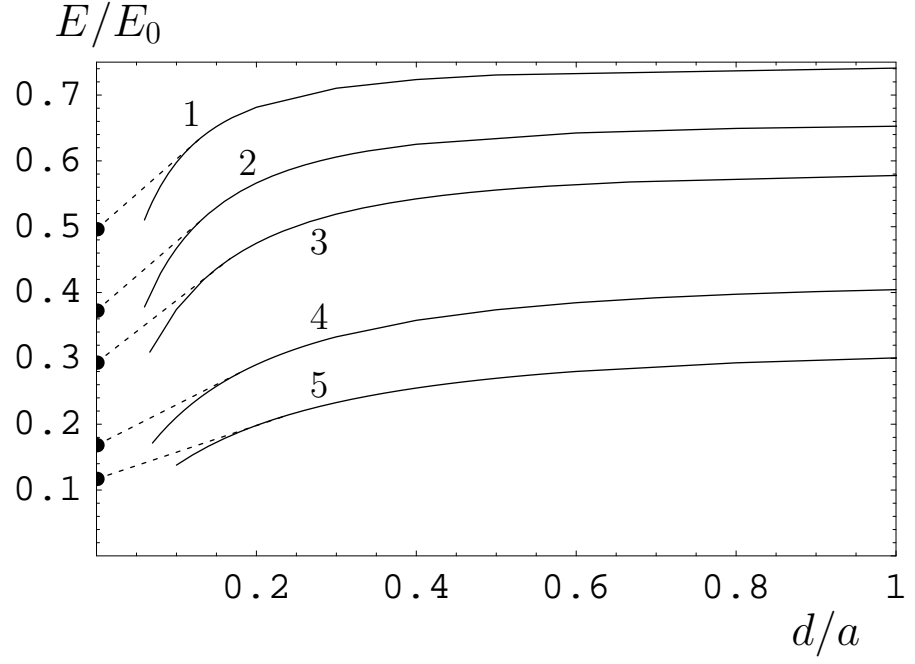}
\vspace*{-10cm}
\caption{The normalized to the case of ideal metals van der Waals
and Casimir energy per unit area between a graphite plate of
thickness $d$ and Au semispace versus relative thickness.
Lines 1,\,2,\,3,\,4, and 5 are related to separations
$a=1,\,0.5,\,0.3,\,0.1,$ and $0.05\,\mu$m, respectively.
The respective normalized interaction energies of a graphene with
an Au semispace are marked on the vertical axis (see text for further
discussion).
}
\end{figure*}
\begin{figure*}
\vspace*{-7cm}
\includegraphics{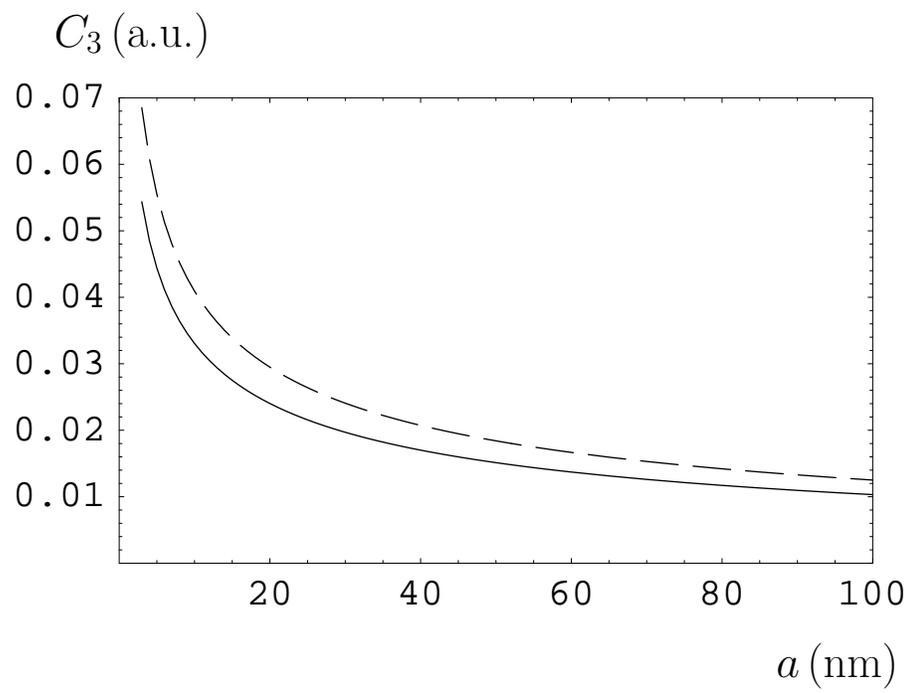}
\vspace*{-10cm}
\caption{The van der Waals coefficient $C_3$ for the interaction
of a hydrogen atom (the solid line) and a molecule (the dashed line)
with graphene versus separation.
}
\end{figure*}
\begin{figure*}
\vspace*{-7cm}
\includegraphics{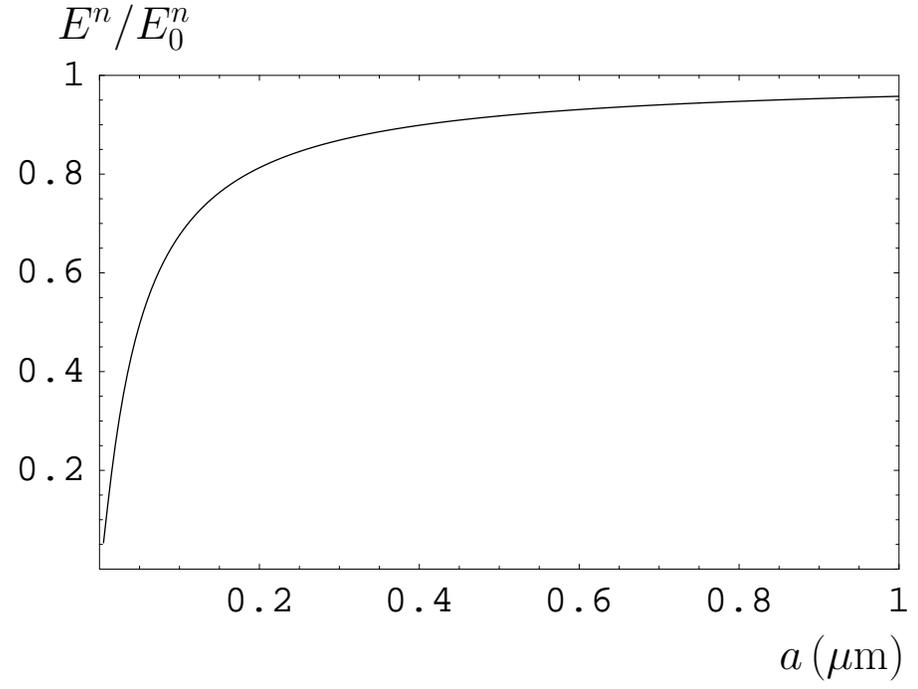}
\vspace*{-10cm}
\caption{The normalized to the case of ideal metals Casimir
energy per unit length between a carbon nanotube and Au semispace
versus separation.
}
\end{figure*}
\begingroup
\squeezetable
\begin{table}
\caption{Ratios $E(a)/E_0(a)$ for the Casimir interaction of
an Au semispace
and graphene computed using: (a) --- tabulated optical
data for Au, (b) ---
plasma model for Au, (c) --- asymptotic expression (\ref{eq23}).
}
\begin{ruledtabular}
\begin{tabular}{clll}
\multicolumn{1}{c}{$a$}&\multicolumn{3}{c}{$E(a)/E_0(a)$} \\
(nm) & $\phantom{X}$(a) &  $\phantom{X}$(b) &  $\phantom{X}$(c) \\
\hline
3 & 0.0252 &0.0222 & 0.0223 \\
5 & 0.0338 & 0.0307 & 0.0310 \\
10 & 0.0496 & 0.0466 & 0.0473 \\
15 & 0.0618 & 0.0590 & 0.0602 \\
20 &0.0720 &0.0694 & 0.0711 \\
30 & 0.0893 & 0.0870 & 0.0895 \\
40 & 0.104 & 0.102 & 0.105 \\
50 & 0.117 & 0.115 & 0.119 \\
60 & 0.129 & 0.127 & 0.131 \\
70 & 0.139 & 0.138 & 0.143 \\
80 & 0.1496 & 0.1485 & 0.154
\end{tabular}
\end{ruledtabular}
\end{table}
\endgroup
\begingroup
\squeezetable
\begin{table}
\caption{Ratios $E^n(a)/E_0^n(a)$ and $F^n(a)/F_0^n(a)$
for the van der Waals interaction of a carbon nanotube with
Au and Si semispaces
}
\begin{ruledtabular}
\begin{tabular}{lllll}
\multicolumn{1}{c}{$a$}&\multicolumn{2}{c}{$E^n(a)/E_0^n(a)$}&
\multicolumn{2}{c}{$F^n(a)/F_0^n(a)$} \\
(nm) & $\phantom{X}$Au &  $\phantom{X}$Si & $\phantom{X}$Au &
$\phantom{X}$Si  \\
\hline
1& 0.0151 &0.0126 &0.0114 &0.00945 \\
1.5& 0.0193 & 0.0162 & 0.0147& 0.0123\\
2& 0.0230 & 0.0193& 0.0175& 0.0147\\
2.5& 0.0262& 0.0221& 0.0201& 0.0169\\
3& 0.0291& 0.0245& 0.0224& 0.0189
\end{tabular}
\end{ruledtabular}
\end{table}
\endgroup

\begin{thebibliography}{99}
\bibitem{0}
J.~Mahanty and B.~W.~Ninham,
{\it Dispersion Forces} (Academic Press, New York, 1976).
\bibitem{01}
H.~B.~G.~Casimir,
{ Proc. K. Ned. Akad. Wet.}
{\bf 51}, 793 (1948).
\bibitem{02}
H.~B.~G.~Casimir and D.~Polder,
{Phys. Rev.}
{\bf 73}, 360 (1948).
\bibitem{1}
R.~Saito, G.~Dresselhaus, and M.~S.~Dresselhaus,
{\it Physical Properties of Carbon Nanotubes}
(Imperial College Press, London, 1998).
\bibitem{2}
A.~C.~Dillon, K.~M.~Jones, T.~A.~Bekkedahl, C.~H.~Kiang,
D.~S.~Bethune, and M.~J.~Heben,
Nature {\bf 386}, 377 (1997).
\bibitem{3}
R.~G.~Ding, G.~Q.~Lu, Z.~F.~Yan, and M.~A.~Wilson,
J. of Nanoscience and Nanotech. {\bf 1}, 7 (2001).
\bibitem{4}
W.~A.~Di\~{n}o, H.~Nakanishi, and H.~Kasai,
e-J. Surf. Sci. Nanotech. {\bf 2}, 77 (2004).
\bibitem{5}
I.~V.~Bondarev and Ph.~Lambin,
Solid State Commun. {\bf 132}, 203 (2004).
\bibitem{6}
I.~V.~Bondarev and Ph.~Lambin,
Phys. Rev. B {\bf 72}, 035451 (2005).
\bibitem{7}
Ch.~Girard, Ph.~Lambin, A.~Dereux, and A.~A.~Lucas,
Phys. Rev. B {\bf 49}, 11425 (1994).
\bibitem{8}
A.~Bogicevic, S.~Ovesson, P.~Hyldgaard, B.~I.~Lundqvist,
H.~Brune, and D.~R.~Jennison,
Phys. Rev. Lett. {\bf 85}, 1910 (2000).
\bibitem{9}
E.~Hult, P.~Hyldgaard, and B.~I.~Lundqvist,
Phys. Rev. B {\bf 64}, 195414 (2001).
\bibitem{10}
H.~Rydberg, M.~Dion, N.~Jacobson, E.~Schr\"{o}der,
 P.~Hyldgaard, S.~I.~Simak, D.~C.~Landreth, and B.~I.~Lundqvist,
Phys. Rev. Lett. {\bf 91}, 126402 (2003).
\bibitem{11}
J.~Jung, P.~Garc\'{\i}a-Gonz\'{a}lez, J.~F.~Dobson, and
R.~W.~Godby,
Phys. Rev. B {\bf 70}, 205107 (2004).
\bibitem{12}
J.~Kleis, P.~Hyldgaard, and E.~Schr\"{o}der,
Comp. Mat. Sci. {\bf 33}, 192 (2005).
\bibitem{13}
J.~F.~Dobson, A.~White, and A.~Rubio,
Phys. Rev. Lett. {\bf 96}, 073201 (2006).
\bibitem{13a}
L.~A.~Girifalco and M.~Hodak,
Phys. Rev. B {\bf 65}, 125404 (2002).
\bibitem{14}
E.~V.~Blagov, G.~L.~Klimchitskaya, and V.~M.~Mostepanenko,
Phys. Rev. B {\bf 71}, 235401 (2005).
\bibitem{15}
E.~M.~Lifshitz,
Zh. Eksp. Teor. Fiz. {\bf 29}, 94 (1956)
[Sov. Phys. JETP  {\bf 2}, 73 (1956)].
\bibitem{16}
E.~M.~Lifshitz and L.~P.~Pitaevskii,
{\it Statistical Physics} (Pergamon Press, Oxford, 1980), Pt.\ II.
\bibitem{17}
M.~Bordag, U.~Mohideen, and V.~M.~Mostepanenko,
{ Phys. Rep.} {\bf 353}, 1 (2001).
\bibitem{18}
F.~Chen, U.~Mohideen, G.~L.~Klimchitskaya, and
V.\ M.\ Mos\-te\-pa\-nen\-ko,
Phys. Rev. Lett. {\bf 88}, 101801 (2002);
Phys. Rev. A {\bf 66}, 032113 (2002).
\bibitem{19}
R. S. Decca, E.
Fischbach, G. L. Klimchitskaya, D. E. Krause, D. L\'opez, and V.
M. Mostepanenko, Phys. Rev. D {\bf 68}, 116003 (2003).
\bibitem{20}
F.~Chen, G.~L.~Klimchitskaya, U.~Mohideen, and
V.\ M.\ Mos\-te\-pa\-nen\-ko,
Phys. Rev. A {\bf 69}, 022117 (2004).
\bibitem{21}
F.~Chen, U.~Mohideen, G.~L.~Klimchitskaya, and
V.~M.~Mos\-te\-pa\-nen\-ko,
{Phys. Rev. A} {\bf 72}, 020101(R) (2005);
{\bf 74}, 022103 (2006).
\bibitem{22}
J.~F.~Babb, G.~L.~Klimchitskaya, and V.~M.~Mostepanenko,
Phys. Rev. A {\bf 70}, 042901 (2004).
\bibitem{23}
M.~Antezza, L.~P.~Pitaevskii, and S.~Stringari,
Phys. Rev. A {\bf 70}, 053619 (2004).
\bibitem{24}
A.~L.~Fetter,
Ann. Phys. (N.Y.) {\bf 81}, 367 (1973).
\bibitem{25}
G.~Barton,
J. Phys. A: Math. Gen. {\bf 37}, 1011 (2004).
\bibitem{26}
G.~Barton,
J. Phys. A: Math. Gen. {\bf 38}, 2997 (2005).
\bibitem{27}
M.~Bordag,
J. Phys. A: Math. Gen. {\bf 39}, 6173 (2006).
\bibitem{28}
G.~A.~Gallup, Chem. Phys. Lett. {\bf 187}, 187 (1991).
\bibitem{29}
A.~O.~Caride, G.~L.~Klimchitskaya, V.~M.~Mostepanenko, and
S.~I.~Zanette,
Phys. Rev. A {\bf 71}, 042901 (2005).
\bibitem{30}
{\it Handbook of Optical Constants of Solids},
ed. E.~D.~Palik (Academic Press, New York, 1985).
\bibitem{31}
A.~Rauber, J.~R.~Klein, M.~W.~Cole, and L.~W.~Bruch,
Surf. Sci. {\bf 123}, 173 (1982).
\bibitem{32}
J.~Blocki, J.~Randrup, W.~J.~Swiatecki, and C.~F.~Tsang,
Ann. Phys. (N.Y.) {\bf 105}, 427 (1977).
\bibitem{33}
M.~Bordag,
Phys. Rev. D {\bf 73}, 125018 (2006).
\bibitem{34}
T.~Emig, R.~L.~Jaffe, M.~Kardar, and A.~Scardicchio,
Phys. Rev. Lett. {\bf 96}, 080403 (2006).
\bibitem{35}
A.~Bulgac, P.~Magierski, and A.~Wirzba,
Phys. Rev. D {\bf 73}, 025007 (2006).
\bibitem{36}
P.~E.~Lammert and V.~H.~Crespi,
Phys. Rev. Lett. {\bf 85}, 5190 (2000);
Phys. Rev. B {\bf 69}, 035406 (2004).
\bibitem{36a}
J.~Israelachvili,
{\it Intermolecular and Surface Forces}
(Academic Press, New York, 1992).
\end{thebibliography}
\end{document}